\documentclass[11pt]{article}
\usepackage[left=1in,top=1in,right=1in,bottom=1in]{geometry}
\usepackage{times}
\usepackage{verbatim}
\usepackage{amsmath}
\usepackage{amssymb}
\usepackage{amsthm}
\usepackage{rotating}
\usepackage{algorithm}
\usepackage[noend]{algpseudocode}

\usepackage{silence}
\usepackage{tabularx}
\usepackage{graphicx}
\usepackage{url}
\usepackage{multirow}
\usepackage{subfig}

\usepackage[noend]{algpseudocode}
\usepackage{siunitx}

\usepackage{array}      % for \extrarowheight macro
\newcommand\mc[1]{\multicolumn{1}{c}{#1}} % handy shortcut macro

\allowdisplaybreaks

\begin{document}
\title{Nonparametric Strategy Test}
\author{Sam Ganzfried\\
Ganzfried Research \\
sam.ganzfried@gmail.com
}

\date{\vspace{-5ex}}

\maketitle

\begin{abstract}
We present a nonparametric statistical test for determining whether an agent is following a given mixed strategy in a repeated strategic-form game given samples of the agent's play. This involves two components: determining whether the agent's frequencies of pure strategies are sufficiently close to the target frequencies, and determining whether the pure strategies selected are independent between different game iterations. Our integrated test involves applying a chi-squared goodness of fit test for the first component and a generalized Wald-Wolfowitz runs test for the second component. The results from both tests are combined using Bonferroni correction to produce a complete test for a given significance level $\alpha.$ We applied the test to publicly available data of human rock-paper-scissors play. The data consists of 50 iterations of play for 500 human players. We test with a null hypothesis that the players are following a uniform random strategy independently at each game iteration. Using a significance level of $\alpha = 0.05$, we conclude that 305 (61\%) of the subjects are following the target strategy.
\end{abstract}

\section{Introduction}
\label{se:intro}
A \emph{strategic-form game} consists of a finite set of players $N = \{1,\ldots,n\}$, a finite set of pure strategies $S_i$ for each player $i \in N$, and a real-valued utility for each player for each strategy vector (aka \emph{strategy profile}), $u_i : \times_i S_i \rightarrow \mathbb{R}$. A \emph{mixed strategy} $\sigma_i$ for player $i$ is a probability distribution over pure strategies, where $\sigma_i(s_{i'})$ is the probability that player $i$ plays pure strategy $s_{i'} \in S_i$ under $\sigma_i$. Let $\Sigma_i$ denote the full set of mixed strategies for player $i$. A strategy profile $\sigma^* = (\sigma^*_1,\ldots,\sigma^*_n)$ is a \emph{Nash equilibrium} if $u_i(\sigma^*_i,\sigma^*_{-i}) \geq u_i(\sigma_i, \sigma^*_{-i})$ for all $\sigma_i \in \Sigma_i$ for all $i \in N$, where $\sigma^*_{-i} \in \Sigma_{-i}$ denotes the vector of the components of strategy $\sigma^*$ for all players excluding player $i$. Here $u_i$ denotes the expected utility for player $i$, and $\Sigma_{-i}$ denotes the set of strategy profiles for all players excluding player $i$. In a repeated game, the original game is repeated some number of iterations and the payoff is the sum over the iterate payoffs. As an example of strategic-form game, the payoff matrix for rock-paper-scissors (RPS) is provided below.

\[
\setlength{\extrarowheight}{2pt}
    \begin{array}{rr|c|c|c|}
        & \mc{}         & \multicolumn{3}{c}{\textup{Player 2}} \\
        & \mc{}         & \mc{R} & \mc{P} & \mc{S} \\ \cline{3-5}
        & R             &  (0,0) & (-1,1) & (1,-1) \\ \cline{3-5}
        \textup{Player 1}
        & P             & (1,-1) &  (0,0) & (-1,1) \\ \cline{3-5}
        & S             & (-1,1) & (1,-1) &  (0,0) \\ \cline{3-5}
    \end{array}
\]

Often the goal of an agent is to play a strategy that corresponds to an established game-theoretic solution concept, such as Nash equilibrium. Nash equilibrium has particularly compelling properties in two-player \emph{zero-sum games}, which are adversarial games where the sum of all payoffs equals zero at each strategy profile. For example, RPS has a unique Nash equilibrium in which each player selects each pure strategy with probability $\frac{1}{3}.$ However, following a Nash equilibrium (or a different solution concept) is not necessarily the goal. When playing against suboptimal opponents who are themselves not playing a Nash equilibrium, we can obtain a significantly higher payoff by learning and exploiting a model of their strategy. 

Typical approaches for opponent modeling and exploitation work by creating a model for an opponent's strategy based on frequency counts observed from samples of play~\cite{Johanson07:Computing,Ganzfried11:Game}. For example, if we observe an opponent play rock 25 times, paper 60 times, and scissors 15 times over 100 iterations of play, then it seems sensible to conclude that they are selecting actions with probabilities (0.25,0.6,0.15). Given this opponent model, we may elect to play scissors with larger frequency as it is a best response to this strategy. Note that this model explicitly assumes that the opponent is following a static mixed strategy (i.e., the action probabilities do not change over time). This model is also implicitly assuming that the opponent is randomizing independently at each iteration. If this assumption fails, we may potentially miss out on a significant amount of expected payoff. For example, suppose the opponent is playing a deterministic strategy that plays rock with probability 1 at iteration 1, paper with probability 1 at iteration 2, scissors with probability 1 at iteration 3, and so on. Then their aggregate frequencies will tend towards $\left(\frac{1}{3}, \frac{1}{3}, \frac{1}{3}\right)$, and so we may decide to follow $\left(\frac{1}{3}, \frac{1}{3}, \frac{1}{3}\right)$ ourselves thinking there is no room for exploitation, while in fact we could actually obtain a payoff of 1 each iteration by correctly predicting and exploiting their strategy.

The Deviation-Based Best Response algorithm works by first following an (approximate) Nash equilibrium strategy for ourselves during the first $E$ iterations, while collecting observations of the opponent's play~\cite{Ganzfried11:Game}. For subsequent iterations we compute and play a best response to the opponent's strategy, while continuing to update our observations. This approach has been demonstrated to obtain significantly higher payoff against weak agents in two-player limit Texas hold'em than following the approximate Nash equilibrium strategy throughout all iterations. However, such an approach is also susceptible to being counterexploited by a strong and/or deceptive opponent. If after the $E$ initial plays have been observed we have not identified any clear weaknesses in the opponent's play, it is likely prudent to continue playing our equilibrium strategy instead of performing potentially risky exploitation. For example, suppose that we are playing a game with a known unique Nash equilibrium $\sigma^*$, and suppose we have a test $\phi$ that determines whether the opponent is following $\sigma^*$ based on our observations of their play so far. We can integrate $\phi$ into our opponent modeling procedure to obtain a new meta-algorithm, depicted in Algorithm~\ref{al:meta}, that only attempts to exploit opponents whose play is sufficiently far from $\sigma^*$. The procedure assumes an arbitrary opponent modeling algorithm $M$ as input (which could just be based on frequency counts). The opponent modeling algorithm takes the observations of play as input, as well as optionally a prior distribution if one is available. We also assume that we are given a response function $R$ as input (which could just be a best response). If after the exploratory phase concludes our statistical test $\phi$ determines that the opponent is following $\sigma^*$, then we do not attempt to exploit and instead prefer to just follow our component of $\sigma^*$ for all remaining timesteps. Otherwise, we proceed to update and respond to our opponent model for subsequent timesteps. 

\begin{algorithm}[!ht]
\caption{Meta-algorithm for opponent modeling and exploitation with a statistical strategy test}
\label{al:meta} 
\textbf{Inputs}: Number of exploration timesteps $E$, total number of timesteps $T$, Nash equilibrium strategy profile $\sigma^*$, prior strategy distribution $p_0$, opponent modeling algorithm $M$, response function $R$, strategy test $\phi$
\begin{algorithmic}
\For {$i$ = 1 to $E$}
\State Play according to $\sigma^*$ and collect observation of opponent's play $O_i$
\EndFor
\State $\hat{\sigma} = M(p_0, (O_1,\ldots,O_E))$
\If {$\phi(\hat{\sigma}, \sigma^*)$ accepts the null hypothesis}
\State Play according to $\sigma^*$ for all remaining iterations
\Else
\State $\tau = R(\hat{\sigma})$
\For {$i = E+1$ to $T$}
\State Play according to strategy $\tau$
\State Collect observation of opponent's play $O_i$
\State $\hat{\sigma} = M(p_0, (O_1,\ldots,O_i))$
\State $\tau = R(\hat{\sigma})$
\EndFor
\EndIf
\end{algorithmic}
\end{algorithm}

Thus, one potential benefit of an accurate statistical test for strategy equivalence is improved opponent modeling algorithms that have reduced risk of being counterexploited by strong and/or deceptive opponents. A related scenario would be if we knew that our opponent was following one of a small number of strategies, but we were unsure of which one. Beyond improved opponent modeling and exploitation, a statistical strategy test could be useful for ensuring the security of a system against certain harmful strategies. For example, the use of bots and/or real-time computer assistance is prohibited in online games such as poker and chess, and certain behavior is also prohibited in many auctions and financial trading platforms. A successful statistical strategy test could ensure that such nefarious behavior is identified so that violators are penalized.

\section{Statistical Test}
\label{se:stat}
In this section we present our nonparametric statistical test for determining whether a player $i$ is following a given mixed strategy $\sigma^*_i.$ Suppose that player $i$ has $k = |S_i|$ pure strategies, and that we have observed a sequence of $n$ plays $O = (O_1,\ldots,O_n),$ where $O_j \in S_i.$ Suppose that $S_i = \{s_1,\ldots,s_k\},$ and that $s_j$ has been observed $n_j$ times in $O$. Our null hypothesis $H_0$ is that player $i$ is following $\sigma^*_i$. Note that this requires both that their pure strategy frequencies are the same as those of $\sigma^*_i$ and also that the actions are selected independently in the different iterations (i.e., that the sequence $O$ is random). The alternative hypothesis $H_a$ is that player $i$ is not following $\sigma^*_i.$

Given a significance level $\alpha$, our test will separately perform a runs test of independence and a chi-squared test for goodness of fit. Each of these tests will be performed with significance level $\frac{\alpha}{2}$ so that they can be combined using Bonferroni correction. Our overall test will reject $H_0$ if at least one of the individual tests rejects at the $\frac{\alpha}{2}$ level. The overall algorithm is given by Algorithm~\ref{al:overall}. The first subroutine (Algorithm~\ref{al:runs}) is a generalized version of the runs test for more than two categories.\footnote{\url{https://real-statistics.com/non-parametric-tests/one-sample-runs-test/runs-test-with-more-than-two-categories/} adapted from~\cite{Sheskin00:Handbook}.}\footnote{Prior work has used the two-category runs test to analyze chimpanzee play in several competitive scenarios~\cite{Martin14:Chimpanzee}.} Recall that a run is a consecutive subsequence of the same category. For example, in the play sequence of RRPPPSPRR of rock-paper-scissors there are five runs. Note that $\Phi$ denotes the cdf of the standard normal distribution. The second subroutine (Algorithm~\ref{al:gf}) performs a chi-squared goodness of fit test comparing the observed counts from $O$ to the expected counts according to the target strategy $\sigma^*_i.$ Algorithm~\ref{al:overall} runs both of these tests separately and rejects the null hypothesis if at least one of the $p$-values is less than or equal to $\frac{\alpha}{2}$. By the Bonferroni correction the full test has significance $\alpha.$ 

\begin{algorithm}[!ht]
\caption{Generalized runs test (GRT)}
\label{al:runs} 
\textbf{Inputs}: Sequence of observations $O = (O_1,\ldots,O_n)$, pure strategy counts $(n_1,\ldots,n_k)$
\begin{algorithmic}
\State $r$ = number of runs in $O$
\State $q = \sum_{j=1}^k n^2_j$
\State $c = \sum_{j=1}^k n^3_j$
\State $\mu = \frac{n(n+1)-q}{n}$
\State $\sigma = \sqrt{\frac{q[q + n(n+1)] - 2nc - n^3}{n^2(n-1)}}$
\State $z = \frac{r - \mu}{\sigma}$
\State \Return $2 \Phi(-|z|)$
\end{algorithmic}
\end{algorithm}

\begin{algorithm}[!ht]
\caption{Chi-squared goodness of fit test ($\chi^2$GOFT)}
\label{al:gf} 
\textbf{Inputs}: Target mixed strategy $\sigma^*_i = (p_1,\ldots, p_k)$, pure strategy counts $(n_1,\ldots,n_k)$
\begin{algorithmic}
\State $T = \sum_{j=1}^k \frac{(n_j - n p_j)^2}{n p_j}$
\State \Return p-value of $T$ for chi-squared distribution with $k-1$ degrees of freedom
\end{algorithmic}
\end{algorithm}

\begin{algorithm}[!ht]
\caption{Overall nonparametric strategy test}
\label{al:overall} 
\textbf{Inputs}: Target mixed strategy $\sigma^*_i = (p_1,\ldots, p_k)$, sequence of observations $O = (O_1,\ldots,O_n)$, desired significance level $\alpha$
\begin{algorithmic}
\State $n_j$ = number of times player $i$'s $j$th pure strategy $s_j$ is observed in $O$
\State $p_r$ = GRT($O$, $(n_1,\ldots,n_k)$)
\State $p_{\chi}$ = $\chi^2$GOFT($\sigma^*_i$, $(n_1,\ldots,n_k)$)
\If {$p_r \leq \frac{\alpha}{2}$ or $p_{\chi} \leq \frac{\alpha}{2}$}
\State Reject the null hypothesis and conclude that the opponent is not playing $\sigma^*_i$
\Else
\State Accept the null hypothesis and conclude that the opponent is playing $\sigma^*_i$ 
\EndIf
\end{algorithmic}
\end{algorithm}

\section{Experiments}
\label{se:exp}
We experimented on public data of human play on rock-paper-scissors.\footnote{\url{https://github.com/kuro-lab/RPSdata}.} The data consists of 50 consecutive plays
of the game from 500 different human subjects. For each subject the data consists of a CSV file which contains 50 elements from $\{0,1,2\}.$ We can associate each of these integers with one of the pure strategies in RPS (though it is not stated which integer corresponds to which action). Our results are depicted in Table~\ref{ta:results}. For $\alpha = 0.05$ and $0.025$, we report how many of the subjects were significant for each combination of the two statistical tests. With $\alpha = 0.025$ our overall test concludes that $500 - 305 = 195$ (39\%) of the subjects are following strategies inconsistent with the target strategy, while 305 (61\%) are following the target strategy. If we had used $\alpha = 0.05$ and just looked at the result of each test separately, we conclude that $500 - (14 + 22) = 464$ subjects are playing strategy frequencies consistent with the target, and $500 - (14 + 196) = 290$ are playing random strategies whose actions are selected independently in different iterations.

\begin{table}[!ht]
\centering
\begin{tabular}{|*{5}{c|}} \hline
$\alpha$ &$X_1$ &$X_2$ &$X_3$ &$X_4$\\ \hline
0.05 &14 &22 &196 &268 \\ \hline
0.025 &10 &17 &168 &305 \\ \hline
\end{tabular}
\caption{Results for 500 human subjects. Column $X_1$ is the number of subjects for which both the chi-squared and runs tests were statistically significant at the $\alpha$ level; column $X_2$ is the number of subjects for which only the chi-squared test was significant; column $X_3$ is the number of subjects for which only the runs test was significant; and column $X_4$ is the number of subjects for which neither test was significant.}
\label{ta:results}
\end{table}

The data is from a study described by Komai et al.~\cite{Komai22:Human}. Note that the subjects are attempting to maximize their payoff against a computer opponent and are not necessarily attempting to follow the target strategy. They conclude that in aggregate the Lempel-Ziv complexity of human play differed from pseudorandom data, though quantitative results were not provided. They also conclude from a recurrence plot that human play is ``less complex and more deterministic'' than pseudorandom data. A prior study on a different dataset has rejected uniform random play for 18\% of ``experienced'' RPS players based on a chi-squared test~\cite{Batzilis19:Behavior}. In comparison, our chi-squared test rejected uniform frequencies of play for 7.2\% of the subjects.

\section{Conclusion}
\label{se:conc}
We presented a nonparametric statistical test that determines whether an opponent is following a specified target mixed strategy in a repeated strategic-form game.
The test ensures both that the opponent's overall frequencies agree with those of the target strategy and also that the actions are chosen independently between iterations. Our experiments on a public dataset of human rock-paper-scissors play show that 61\% of the subjects are following the target strategy (independently selecting each action with probability $\frac{1}{3}$). Our test has applications to improved opponent modeling approaches that obtain higher payoff against suboptimal opponents and improved security for detection of malicious agents. Our test is efficient and scalable to large games. We leave the problem of extending the approach to imperfect-information games for future study.

\bibliographystyle{plain}
\bibliography{C://FromBackup/Research/refs/dairefs}

\end{document}